\documentclass{elsarticle}
\makeatletter
\def\ps@pprintTitle{%
	\let\@oddhead\@empty
	\let\@evenhead\@empty
	\let\@oddfoot\@empty
	\let\@evenfoot\@oddfoot
}
\makeatother
\usepackage{lineno,hyperref}
\modulolinenumbers[5]
\usepackage{setspace}
\usepackage{amsmath}
\usepackage{amssymb}
\usepackage{braket}
\DeclareMathAlphabet{\mathpzc}{OT1}{pzc}{m}{it}

\begin{document}

\begin{frontmatter}

\title{Deconstructing {\" U}nsal-Yaffe Reconfinement.}

\author{Herbert Neuberger}\fnref{myemail}
\fntext[myemail]{Email: herbert.neuberger@gmail.com}

\address{Department of Physics and Astronomy, Rutgers University,  Piscatway, NJ 08854 U.S.A.}

\begin{abstract} In the UY reconfined phase on a lattice the thermal trace is over states transforming in all $SU(N)/Z(N)$ irreducible representations as opposed to only over $SU(N)$ singlets in the standard formulation.  As $N\to\infty$, on a finite lattice, the usual Hilbert space becomes orthogonal to the deformed one. Concerns about the extended UY proposal for large $N$ Eguchi-Kawai reduction are raised.
 \end{abstract}

\begin{keyword}
\texttt{ Lattice Gauge Theory, Finite Temperature, Large N.}
\end{keyword}

\end{frontmatter}

\section{Introduction.}

In~\cite{Unsal:2008ch} {\" U}nsal and Yaffe (UY) proposed a deformation of pure $SU(N)$ finite temperature (T) gauge theory which replaces its deconfined phase by a ``reconfined'' one. Subsequently, workers in lattice gauge theory concluded that this phase is similar to ordinary $SU(N)$ for several observables ~\cite{Bonati:2020lal,Athenodorou:2020clr}. It was also suggested by UY that extending their idea to four directions would produce ordinary planar gauge theory on an $1^4$ lattice, at $N\to\infty$, removing a flaw in the original EK model~\cite{Eguchi:1982nm}. In ~\cite{Bhanot:1982sh} a quenched alteration, QEK, was early on conjectured as a fix. A suggestion for repairing EK similar to that of UY  followed it not too long thereafter~\cite{Chen:1982uz}.

Bringoltz and Sharpe (BS)~\cite{Bringoltz:2008av} revisited QEK, finding that it fails, and,  quoting a purported failure~\cite{Azeyanagi:2007su,TEPER2007359} of the TEK (Twisted Eguchi-Kawai) model ~\cite{Gonzalez-Arroyo:1982hwr}, concluded that only DEK (Deformed Eguchi-Kawai) remained viable. In ~\cite{Neuberger:2020wpx} it was shown that the issue raised by BS had to do with a faulty change of variables and disappeared when it was done correctly. Similarly, TEK has been shown to survive its criticism: for a review see ~\cite{GarciaPerez:2020gnf}.  At the present, I am skeptical about DEK being right. 

QEK and TEK are two methods that employ drastic changes, while the UY deformation superficially seems innocuous. I shall show that the finite T lattice models, as simulated, are thermodynamically different from normal gauge theory in a dramatic way. Consequently, there is reason of concern also about DEK. 

This is the main point of the present paper. In this work there will be no rigorous proof that DEK fails to reproduce ordinary T=0 pure gauge theory in the planar limit and infinite space. 

In general, I believe that the breakdown of global $Z(N)$'s is a useful symptom of failures in attempts at maximal reduction, but not the root cause to be addressed by a credible cure. The problem is completely framed within a perfectly defined integral over four $SU(N)$ matrices as $N\to \infty$. EK reduction to an $1^4$ lattice seems a miracle because there is no room for Faraday's lines to spread and nevertheless Coulomb's law has to hold. A believable cure requires a direct and clear mechanism for how this ``swindle''~\cite{polchinski1992strings} works; if it does, the symptom of $Z(N)^4$ breaking disappears as a byproduct.

The EK proof relies on preservation of the $Z(N)^4$. Its spontaneous breaking at $N=\infty$ indicates a failure of the proof in some regime. The breaking  is reflected in the distribution of angles $\theta^\mu_j$, $j=1,..,N$  where  $e^{i\theta^\mu_j}$'s are eigenvalues of the $\mu^{\rm th} SU(N)$ matrix in EK. A sort of order parameter, $P_\mu$, was introduced in equation (18)~\cite{Bhanot:1982sh}, for numerical purposes. 
\begin{equation}	
	P_\mu = \frac{1}{N^2} \sum_{i,j} \sin^2 \frac{1}{2} (\theta^\mu_i-\theta^\mu_j)\le \frac{1}{2}	
\end{equation}
The upper bound is saturated if the angles' distribution is invariant under $\frac{2\pi}{N}$ shifts preserving $Z(N)$. Hence, if $\frac{1}{2} > P_\mu$, $Z(N)$ is broken. Conversely, $P_\mu =\frac{1}{2}$ fails to imply $Z(N)$ invariance. For example, when $N$ is factorizable the preservation of only a proper $Z(N)$ subgroup also gives  $P_\mu=\frac{1}{2}$. Taking $N\to\infty$ over primes avoids this at some computational cost, a lazy avoidance strategy I employed in my numerical work. 

\section{Deconstruction.}

UY discuss the physics of finite T as an effective theory. I shall address this part of their work only very briefly later on. In order to avoid dealing with the local - nonlocal combination of terms in the continuum version  I choose to apply deconstruction ~\cite{Arkani-Hamed:2001nha}  and replace the fourth direction by a finite periodic lattice. This theory stays purely three dimensional in the UV and also brings the model closer to lattice simulations.

The deformed theory can be fundamentally different from the undeformed one, consistently with the EFT picture of~\cite{Unsal:2008ch}, but not quite with the numerical results~\cite{Bonati:2020lal,Athenodorou:2020clr}. Also, the DEK cure becomes  implausible. DEK would have to reproduce usual pure $SU(N)$ gauge theory at $N=\infty$, on the lattice, and at finite volume, \underline{exactly} -- not just in any approximate manner.

\section{Finite or high T.}
\subsection{Deformed Hilbert space ${\cal H}_{\rm def}$ and its fate at $N=\infty$.}
Kogut and Susskind~\cite{Kogut:1974ag} wrote down the Hamiltonian $H$ for lattice gauge theory in continuous time.
For a finite spacial volume $H$ acts on the Hilbert ${\cal H}$ space of wave functions  $\Psi(U)$ where $U$ is the collection of link variables on a three dimensional hypercubic lattice. $H$ commutes with local gauge transformations $G(g)$ where $g({\vec x})\in SU(N)$ and ${\vec x}$ is any spacial site, corresponding to time ($t$) independent gauge transformations of link variables perpendicular to $t$, after choosing axial gauge along $t$. ${\cal H}$ can be decomposed into  irreducible representations (irreps) of $G$ given by a collection of independent irreps $\forall {\vec x}$. The Gibbs trace expresses the finite T partition function  (${\cal Z}$) by summing over all states in ${\cal H}$ which are pure singlets under $G$. The projector onto ${\cal H}_{\rm phys}\subset{\cal H} $ is $\mathfrak{P}$:
\begin{equation}\label{eqn:proj}
	(\mathfrak{P} \Psi )(U)=\int [dg] \Psi (^{g} U)
\end{equation}
$^{g}\!U$ is $U$ gauge transformed by $g$ and $[dg]$ is Haar. At finite lattice spacing in the time direction $H$ is replaced by the transfer matrix $\mathfrak{T}=e^{-H}$ where $H\ge 0$. An explicit expression is needed only for $\mathfrak{T}$. To get ${\cal Z}$, $\mathfrak{P}$ must be inserted in ${\rm Tr}[{\mathfrak{T}}^{N_t} ]$. $N_t\ge 2$ is the discrete $\frac{1}{T}$ ($k_B =1$). Canonical and path integral formulations in lattice gauge theory are equivalent.

With the deformation, $[dg]$ is multiplied by a $g$-dependent weight. This replaces equation~(\ref{eqn:proj}) by 
$\mathfrak{Y}$:
\begin{equation}
	\mathfrak{Y} \Psi (U)  = \int [dg] 	e^{\mathfrak{U}(g)} \Psi(^{g}U),
\end{equation}
where $\mathfrak{U}(g)$ is the $Z(N)$ invariant UY deformation term. 

$\mathfrak{Y}$ is not a projector and gets inserted into the thermal trace exactly once.
As an example, I calculate $\mathfrak{Y} \Psi_{ij}$ with $i\ne j$ fixed. $\Psi_{ij}$ is one element of the adjoint, given by an open-corner plaquette at space site ${\vec x}$. It is obvious that $\mathfrak{P}\Psi_{ij}=0$ and the state won't contribute to the thermal trace.  UY add $N^2 h P_4$ to the Wilson action. $\mathfrak{Y}$ differs from $\mathfrak{P}$ at order $h$. One finds:
\begin{equation}
	\mathfrak{Y} \Psi_{ij} \approx h \int dg_{\vec x} |g_{{\vec x}; i,i}|^2 |g_{{\vec x}; j,j}|^2 \Psi_{ij} +{\cal O} (h^2 )
\end{equation}
$\Psi_{ij}$, rather than being annihilated is now multiplied by a constant and contributes to the trace; any $Z(N)$ invariant state also would. For $h\ne 0$ 
an infinite number of such states contribute to the trace, in addition to the singlet sector. 

Canonical quantization of the deformed theory comes with ${\cal H}_{\rm def}$ where ${\cal H}_{\rm phys}\subset{\cal H}_{\rm def} \subset {\cal H}$ and Gauss' law is violated in a major way~\footnote{The equations for $\mathfrak{P}$ and $\mathfrak{Y}$ are easily derived. Maximal axial gauge fixing leaves free one $g_{\vec x}$, connecting two adjacent spacial slices and sets all other to unity. These $g_{\vec x}$ are temporal loops entering the Wilson action through plaquettes with one spacial link in the top slice and a staple based on its parallel link, $U$, in its bottom slice. The entire staple can be replaced in the action just by ${}^g U$. Usually, integration over the $g_{\vec x}$'s is with Haar measures, while the UY deformation adds a $g$--dependent weight factor.}. 

To see this let us calculate ${\rm Tr} [\mathfrak{P}\mathfrak{Y}]$, the trace of $\mathfrak{Y}$ in the singlet sector. Acting there, $\mathfrak{Y}$ just multiplies each state by a constant given by the one link integral:
\begin{equation}
	{\cal J}(2h)\equiv\prod_{i=1}^N [\int_{-\pi \le \theta^4_i <\pi} \frac{d\theta^4_i}{2\pi} ]
	{\prod_{i<j}} [ e^{2h \cos(\theta^4_i - \theta^4_j)}\sin^2\frac{\theta^4_i-\theta^4_j}{2}]
\end{equation}
Expanding in Fourier series one finds:
\begin{equation}
	\frac{{\cal J}(2h)}{{\cal J}(0)}=[ I_0 (2h)-I_1(2h)]^{\frac{N(N-1)}{2}},
\end{equation}
where the $I_n (x)$ are modified Bessel functions. For all $h>0$, $1>I_0(2h)-I_1(2h) > 0$, while  $I_0(2h)\!-\!I_1(2h) > 1$ for all $h<0$. So, ${\cal{J}}(2h) >0$ for all $h$.
At $N=\infty$, $\frac{{\cal J}(2h)}{{\cal J}(0)}$ is $+\infty$ when $h<0$ and $0$ at $h>0$. Hence, as $N\to\infty$,  ${\cal H}_{\rm phys} $ becomes a thermodynamically negligible subset of ${\cal H}_{\rm def}$ at $h<0$. There is one such factor for each spacial site and $N=\infty$ orthogonality holds at a finite number of sites.

\subsection{UY reconfinement versus high T quenching.}

EK failure and high T deconfinement~\cite{Neuberger:1983xc} are somewhat similar. The usual $\theta^4$ dependent effective potential $V_{\rm eff} (\theta^4 )$ to one loop is given by :
\begin{equation}\label{eqn:veff}
	 \sum_{1\le i < j\le N} {\mathfrak{e}} (\theta^4_j -\theta^4_i ),\;
	{\mathfrak{e}}(\gamma )=\frac{2}{L^3} \sum_p \log [1-\frac{\cos\gamma}{T_{N_t} (1+\rho\sum_{\mu=1}^{3}\sin^2\frac{\pi p_\mu}{L})}],
\end{equation}
where $T_{N_t}$ is the Chebyshev polynomial of order $N_t$ and $p_\mu=0,...,L-1$. 
The three volume is $L^3$. $T_{N_t}(1+x^2)$ increases rapidly with $x^2$ since its argument is outside $[-1,1]$, so the relevant qualitative properties for reasonable values of $N_t$ and $\cos \gamma$ obtain already at leading order in the logarithm's expansion. $\rho$ equals the  ratio between couplings of time like and spacial plaquettes. Terms  suppressed by powers of $L$ are neglected. Equation(\ref{eqn:veff}) is an improved version over ~\cite{Neuberger:1983xc} and some typos are fixed. Its derivation in 
~\cite{Neuberger:1983xc} shows explicitly how the quenching ``swindle" restores $T=0$ Feynman diagrams at infinite $N$ and preserves $Z(N)$.

$\gamma=0$ is a minimum of $\mathfrak{e}(\gamma)$ showing that the angles $\theta^4_i$ now have a window of attraction which could break $Z(N)$. 
The probability for coalescing angles is zero; one will always have $\sim\frac{1}{N}$ narrow peaks  housing separated eigenvalues. The cancellation between an approximate one loop result and the exact zero in the measure is only  illusory. $Z(N)$ breaking at $N=\infty$ presents as a modulation of peaks' heights away from uniformity. As $L\to\infty$ we get:
\begin{equation}
	{\mathfrak{e}}^{L=\infty} (\gamma )=2 \int_{|k_\mu|<\pi}\frac{d^3 k}{(2\pi)^3} \log [1-\frac{\cos\gamma}{T_{N_t} (1+\rho\sum_{\mu=1}^{3}\sin^2\frac{k_\mu^2}{2})}]
\end{equation}
Since $N_t \ge 2$ the spacial lattice spacing (now set to 1) can be taken to zero.

UY reconfinement is implemented by a term $\sum_{\vec x} P_4 ({\vec x})$ added to the Wilson action in a way favoring $P_4=1/2$. Each 
$P_4 ({\vec x})$ depends on $e^{i\theta^4_j ({\vec x})}$, the eigenvalues 
of straight Polyakov loops anchored at spacial sites ${\vec x}$.  

One could replace the UY deformation by quenching the $\theta^4_j ({\vec x})$ angles. Then, Matsubara frequencies become a continuum at infinite $N$ and one gets $T=0$
propagators. 

At high T, infinite space and large $N$ the $Z(N)$ breaks spontaneously. This is a symptom of deconfinement, but not the dynamical reason for it: normally, ``continuum smeared''  ~\cite{Narayanan:2006rf} medium sized ($\sim 1 \; {\rm GeV}^{-1}$)  in four dimensions) \underline{contractible} Wilson loops (on which $Z(N)$ has no impact) display eigenvalue distributions that undergo a subtle, operator dependent, phase transition at $N=\infty$ when enlarged, signaling a crossover into an IR dominated regime~\cite{Lohmayer:2011nq}, where the reign of Feynman diagrams ends, at scales preceding confinement, heralding the strong coupling regime of larger distances.  
It is implausible that this is directly caused by a spontaneous breakdown of a global $Z(N)$, not directly coupled to contractible loops. 

More likely, deconfinement reflects some form of intrinsic disorder. Over the years, a menagerie of special configurations that are dynamically responsible for confinement have been suggested. This activity has produced more papers than convincing results. 
Based on~\cite{Lohmayer:2011nq}, I guess that $N=\infty$ confinement might just be some self generated noise by a multitude of different excitations inducing a type of generic disorder and this explains why a random matrix model appears to work. My view is motivated by~\cite{Blaizot2008LargeNcCA} and ~\cite{Shuryak:1992pi}. A presentation at colloquium level is available in ~\cite{Neuberger:2010bq}.

I skip discussing a single Polyakov loop at broken $Z(N)$ indicating finite single quark free energy: this topic was already reviewed by Kiskis~\cite{Kiskis:1994dd}.

\section{High T reconfinement -- yes, EK reduction -- no.}
\subsection{A simple view of reconfinement.}
The standard transition to a plasma at high T has been understood for 
finite and infinite $N$ for a long time~\cite{Hagedorn:1965st,Thorn:1980iv}. In ${\cal H}_{\rm phys}$ the typical lattice states are closed spacial loops of total length $l$.
For low T, large $l$ states are disfavored because confinement makes them heavy. When T is high, the suppression is overwhelmed by the exponentially large number of states 
with large $l$. This entropy dominated situation has so many meandering closed loops 
of essentially infinite length that confinement looses its meaning. 

The UY deformed system has so many states in high representations of $SU(N)/Z(N)$ that they thermally crowd out the long closed loops even at high T. These states are small and confinement of $Z(N)$ charge carrying sources remains meaningful. This argument works for $N_t \ge 2$ but not for $N_t=1$. In the latter case there is Higgs type matter, the system is purely three dimensional, and confinement may come from semiclassical configurations. I am skeptical about this being the single source of disorder. The findings of~\cite{Lohmayer:2011nq} extend to three dimensions~\cite{Narayanan:2007dv}, so likely there also are other sources of disorder. 
\subsection{Large $N$ reduction to $1^4$ by DEK.}

For non-prime $N$ and because of other worries, UY added generalizations of $P_\mu$ to their deformation. Very roughly, the idea is similar to deforming the Ising model with spins $s({\vec x})$ by adding the total magnetization squared, $M^2$, ($M=\sum_{\vec x} s({\vec x})$) to the action, favoring $M=0$ even at strong ferromagnetic coupling. $M=0$ would reflect a random distribution of domains with positive and negative $M$. The same applies to~\cite{Chen:1982uz}. Whether the $Z_2$ symmetry of the Ising model is broken or not, the system remains gapped away from the transition. There is no $N$ in this rough analogue and spontaneous breaking happens only at infinite volume. 

Too often a direct prevention of global $Z(N)^4$ spontaneous breakdown is taken as a mechanism forcing confinement.  This is not the main motivation in QEK and TEK. Yes, $Z(N)^4$'s are kept unbroken, but the essential mechanism is intended to restore a continuous spectrum for translations in flat space (QEK) or, more daringly, takes a route through noncommutative space (TEK). In either way it is evident how propagators in Feynman diagrams get fleshed out to carrying all continuous lattice momenta. Without clear evidence for correct Feynman diagrams, a proposed cure for EK will remain unconvincing. For full large $N$ reduction to work, one requires emergence of non-dynamical, flat, space time from the available dynamical material consisting of $SU(N)$ valued variables. One may view the quenched angles acting as momenta in propagators as an ``emerging'' Euclidean space-time. 

There exists no real proof that loop equations determine usual planar lattice Feynman diagrams. It is not clear what initial conditions need to be added. At strong coupling there is a proof. The relevance of $Z(N)$ symmetry to infinite $N$ lattice loops equations at strong coupling not withstanding, targeting $Z(N)^4$ restoration directly is insufficient -- in my opinion.

The interaction between singlets and non-singlets is suppressed at large $N$. But, it still facilitates thermalization. Large $N$ suppression might be misleading in a non-local framework. For example, one set of exactly massless fermions on the lattice can be ignored at leading order in $N$ only at zero topology. A single instanton makes the path integral vanish. An effect with relative action $-\frac{\infty}{N}$ is not negligible when the $\infty$ is there for all finite $N$. 

\section{Suggestions for numerical tests.}
It would be interesting to simulate the quenched finite T model and compare the results to those of UY reconfinement. The quantum average over $\theta^4({\vec x})$ by UY is replaced by a quenching--forced uniform coverage of $[-\pi, \pi] $ for each ${\vec x}$. 

It would be useful to numerically estimate the thermal mass of adjoint states at UY deformed finite T from the appropriate two point connected correlation function constructed from two relatively $t$-translated open-corner plaquettes (the $\Psi_{ij}$ states in the simple example discussed earlier) with opposing orientations, and connected by straight time like links running up and down on the same path. Summing over all  $i,j$ the correlation function of two $(N^2-1)$--plets $\Psi_{\rm adj}\equiv\Psi_{ij}-\frac{{\rm Tr} \Psi}{N}\delta_{ij}$ gives the $\Psi_{ij}$ mass. 

I predict that in the deformed reconfined regime the $\Psi_{ij}$ has 
a finite mass which significantly depends on $h$ as a consequence of the intrinsic adjoint states  roaming in the environment. This should be seen directly in raw data, after standard decay fits, but with no  extrapolation to continuum and infinite spacial volume, as they are irrelevant for EK reduction. Equilibration of simulations should be carefully monitored, in view of non-locality. For settling the issue one needs results for much higher $N$-values (perhaps prime). 

{\bf Note:} More details are available at \href{URL}{http://arxiv.org/abs/2106.12635}.

\section*{Acknowledgment.}
Support under Rutgers ICR grant 300768 is acknowledged.

\bibliography{herbert.bib}

\end{document}